\documentclass[num-refs]{wiley-article}

\usepackage{graphicx}
\usepackage[space]{grffile}
\usepackage{latexsym}
\usepackage{textcomp}
\usepackage{longtable}
\usepackage{tabulary}
\usepackage{booktabs,array,multirow}
\usepackage{amsfonts,amsmath,amssymb}
\usepackage{natbib}
\usepackage{url}
\usepackage{hyperref}
\hypersetup{colorlinks=false,pdfborder={0 0 0}}
\usepackage{etoolbox}
\newif\iflatexml\latexmlfalse

\AtBeginDocument{\DeclareGraphicsExtensions{.pdf,.PDF,.eps,.EPS,.png,.PNG,.tif,.TIF,.jpg,.JPG,.jpeg,.JPEG}}

\usepackage[utf8]{inputenc}
\usepackage[english]{babel}

\usepackage{siunitx}

\usepackage{color}

\iflatexml

\else
\corraddress{Ma\l gorzata Olejniczak, Centre of New Technologies, University of Warsaw, S. Banacha 2c, 02-097 Warsaw, Poland}
\corremail{malgorzata.olejniczak@cent.uw.edu.pl}
\fundinginfo{
The European Regional Development Fund, projects No. ITMS2014+: 313011W085,
The National Science Centre
(Poland) grant 2016/21/B/ST4/03904,
the National Science Centre
(Poland) grant 2016/23/D/ST4/03217.
}

\fi

\papertype{Full Paper}

\title{Relativistic Frozen Density Embedding calculations of solvent effects on the NMR shielding constants of transition metal nuclei.}

\author[1]{Ma\l gorzata Olejniczak}
\affil[1]{Centre of New Technologies, University of Warsaw, S. Banacha 2c, 02-097 Warsaw, Poland. malgorzata.olejniczak@cent.uw.edu.pl}

\author[2]{Andrej~Antu\v{s}ek}
\affil[2]{ATRI, Faculty of Materials Science and Technology in Trnava,
Slovak University of Technology in Bratislava, J. Bottu 25, 917 24 Trnava \\Slovak Republic. andrej.antusek@stuba.sk}

\author[3]{Micha\l ~Jaszu\'nski}  
\affil[3]{
Institute of Organic Chemistry,
Polish Academy of Sciences
\\Kasprzaka 44/52, 01-224 Warszawa, Poland. michal.jaszunski@icho.edu.pl}

\begin{document}

\maketitle

\begin{abstract}
Nuclear Magnetic Resonance (NMR) shielding constants 
of transition metals in solvated complexes
are computed at the  
relativistic density functional theory (DFT) level. The solvent effects evaluated with subsystem-DFT
approaches are  compared with the 
reference 
solvent shifts predicted from supermolecular calculations.
Two subsystem-DFT approaches are analyzed -- in the 
standard 
frozen density embedding (FDE) scheme the transition metal complexes are embedded in an environment of solvent molecules whose density is kept frozen, in the second approach the 
densities of the complex and of its environment are relaxed in the ``freeze-and-thaw'' 
procedure.
The latter approach 
improves the description of the solvent effects in most cases, nevertheless the FDE deficiencies are rather large in some cases.

\textbf{Keywords} --- Frozen Density Embedding, NMR shielding constant, solvent shifts, transition-metal complexes
\end{abstract}

\section{Introduction}

When we consider heavy atom NMR shielding constants, there are two factors which complicate a reliable comparison 
of the theoretical results with the experimental data. First, the 
relativistic effects 
have to be taken into account, they are too large to be 
determined as corrections to nonrelativistic 
values.\cite{nmr-book-kaupp,jajpta372,mkaupp:rev,jautschbach-rel,aamj-kjmjbook,mr-rel-kjmjbook}
Secondly, gas-phase NMR data is rare for heavy nuclei; only condensed phase spectra are
usually accessible and therefore the effect of the environment should be 
accounted for in the theoretical analysis.
While the  first problem can be 
bypassed by
a combination of
two-component and four-component relativistic Hamiltonians
with the density functional theory (DFT),\cite{jajpta372,mr-rel-kjmjbook}
the latter difficulty 
still remains a challenge.

The modeling of a total system -- consisting of solute and solvent molecules 
-- by quantum chemistry methods is an explicit path to study solvent effects, yet it may 
be unaffordable 
if the number of solvent molecules is large. On the other hand, 
truncating the system to a smaller model or treating the solvent implicitly, for instance as a continuous medium 
like in PCM\cite{jtbmrccr105,diremigio-jpca-119-5061-2015,fortunelli-cpl-231-34-1994,diremigio-ijqc-119-e25685-2018} or COSMO\cite{akgsjcspt2} methods may not lead to sufficiently accurate results,
especially when specific solvent effects occur, such as hydrogen bonds.\cite{jacob-jcp-125-194104-2006}
A frequent practice in these cases is to include few solvent molecules in 
the part of the system treated explicitly by quantum chemistry methods.\cite{mennucci-jpca-105-7287-2001,csjmhojkjctc10} This approach leads to more accurate results, 
but it is computationally cumbersome and cannot be easily generalized.\cite{doi:10.1021/acs.jctc.6b00965}

A promising alternative are subsystem-DFT-based techniques in which the total molecular system is partitioned into subsystems, each handled by the best-suited quantum chemistry model.\cite{senatore-prb-34-5754-1986,cortona-prb-44-8454-1991,wesolowski-jpc-97-8050-1993}
In this work we employ the Frozen Density Embedding (FDE) scheme,\cite{wesolowski-jpc-97-8050-1993} which can be considered as a practical prescription to implement the subsystem-DFT concept.\cite{krishtal-jpcm-27-183202-2015}
FDE is 
based on the partitioning of the total system into 
subsystems, which are treated separately (the computational procedure is set for one subsystem at a time) with 
the interaction between 
subsystems accounted for through the so-called embedding potential. As such, it is an appealing compromise between the (usually costly) calculations on the full system and the calculations on a selected part of the system with its environment either completely or partly neglected, or treated by approximate schemes.
This partitioning has originally been performed in terms of the electron density, however it can also be applied to the current density and to the spin density, what allows to extend the FDE formalism for instance to magnetic properties in non-relativistic~\cite{jacob-jcp-125-194104-2006,bulo-jpca-112-2640-2008} and relativistic contexts~\cite{gotz-jcp-140-104107-2014,morbaspgpccp19}. 
Among many successful applications of FDE,~\cite{krishtal-jpcm-27-183202-2015,
gomes-arc-108-222-2012,jacob-wire-4-325-2014,wesolowski-chemrev-115-5891-2015,sun-acr-49-2705-2016,jones-jacs-142-3281-2020}
 so far
only few studies devoted to 
NMR shielding constants of nuclei in molecules in various environments are reported.\cite{jacob-jcp-125-194104-2006,bulo-jpca-112-2640-2008,halbert-ijqc-2020}
Moreover, the use of four-component relativistic framework in this context was discussed for a few selected systems only.\cite{morbaspgpccp19,halbert-ijqc-2020}

The aim of this work is to test the performance of the FDE method for
modeling solvent effects 
on
NMR shielding constants 
in transition metal complexes used as NMR
standards.\cite{rkhedbsmcmrgpgmric40}
We compare the FDE results with the reference solvent effects estimated from calculations on the total systems (and treated as accurate) and with solvent effects obtained using
other methods.\cite{aamrpccp22}
The selected set of systems represents
an excellent test set for the FDE method:
it contains complexes which are very distinct chemically, exhibit different charge
states, ranging from cations and neutral complexes to anions,
and are in various environments of polar or non-polar solvents
(including e.g. water, acetonitrile and benzene). 
To our best knowledge, it is the first systematic study based on the four-component relatvistic FDE framework of solvent effects on the NMR shielding constants in a large group of complexes, in which relativistic effects are expected to be of various importance.

\section{Theory and computational aspects}

\subsection{FDE scheme - an overview}

In the FDE scheme adapted for the purpose of calculating the NMR shielding 
constant
in the  
relativistic formalism,\cite{morbaspgpccp19} the partitioning of the total system into 
subsystems is realized in terms of the general density component,\cite{bast-ijqc-109-2091-2009,komorovsky-jcp-128-104101-2008,olejniczak-jcp-136-014108-2012}
$\rho_k$ $(k = 0,x,y,z)$
which collects the charge density and the spin-magnetization vector whose norm defines the spin density (non-collinear definition):
\begin{equation}
\rho_{k;\text{super}} = \rho_k^\text{I} + \rho_k^\text{II} 
\label{eq:rho_partitioning}
\end{equation}
In the following 
discussion of FDE, 
supermolecule (denoted as ``super") refers to the total molecular system. 
Each molecular system that we consider can be partitioned  
without loss of generality 
into two subsystems: 
an ``active" one (in our case the transition metal complex), 
which is the moiety with the nucleus X of interest 
and the ``environment" (in our case the solvent).

In the approach adopted here, we treat 
both subsystems separately, with
the geometry of the active subsystem  either
optimized as if in vacuum (``vac")
or with the structures of both subsystems fixed, 
as parts of the optimized geometry of the supermolecule.
The former setting allows to estimate the truly non-interacting limit of properties of the active subsystem,
while in the latter case the supermolecule is partitioned into
subsystems, but their geometries are not relaxed (``isol"). 
Capturing this effect of the environment on the properties of the active subsystem is referred to 
as ``mechanical embedding" in the literature.\cite{gomes-arc-108-222-2012}
FDE adds another layer of environmental effects, as described below. 

The partitioning of the density
 introduced in Eq.~\ref{eq:rho_partitioning} implies that the energy of the total system is expressed as a sum of energies calculated for isolated subsystems and an interaction energy term,
 which depends on the densities of both subsystems
 \begin{equation}
  E_\text{super}[\rho_{k;\text{super}}] = E^\text{I}[\rho_k^\text{I}] + E^\text{II}[\rho_k^\text{II}] + E^\text{(int)}[\rho_k^\text{I}, \rho_k^\text{II}],
  \label{eq:E_partitioning}
\end{equation}
The density of a selected subsystem is then optimized in the 
relativistic 
Kohn-Sham (KS) equations formulated for a constrained electron density (KSCED),\cite{wesolowski-cpl-248-71-1996} in which the KS potential is augmented by a potential accounting for the presence of other subsystems.

This additional \emph{embedding potential}, determined as a functional derivative 
of the interaction energy $E^\text{(int)}[\rho_k^\text{I}, \rho_k^\text{II}]$ with respect to the density of a subsystem 
for which the KSCED equations are set up, depends also on the frozen density of 
the other subsystem, 
therefore it holds the information on the coupling between the subsystems.
The embedding potential involves classical terms (nuclear attraction and Coulomb repulsion of electrons) and non-classical terms that arise as functional derivatives of the non-additive contributions to the exchange-correlation and kinetic energy.\cite{wesolowski-jpc-97-8050-1993}

\paragraph{FDE and the NMR shielding tensor}
It is then
straightforward to adapt the FDE scheme to the response theory. 
The static response equations are derived considering the energy derivatives, 
thus the only additional difficulty in the FDE formulation is due to the need to determine various derivatives of the interaction energy and -- consequently -- of the embedding potential.\cite{casida-ijqc-96-577-2004,neugebauer-jcp-126-134116-2007,neugebauer-jcp-131-084104-2009,hofener-jcp-136-044104-2012}
Here we focus directly on the formulation of the NMR shielding tensor.
As a second-order derivative of the total energy with respect to the external magnetic field (\textbf{B}) and 
the magnetic moment (\textbf{m}$^\text{X}$) of the nucleus X as perturbations, 
in the FDE scheme 
this tensor 
(in the exact case 
equal to the supermolecular property)
can be presented as a sum of contributions from isolated subsystems supplemented by an interaction-related  part 
\begin{equation}
  \boldsymbol\sigma_\text{super}(\text{X})  = 
  \left. \frac{d^2E_\text{super}}{d\textbf{B} d\textbf{m}^\text{X}} \right|_0
  =
  \left. \frac{d^2E^\text{I}}{d\textbf{B} d\textbf{m}^\text{X}} \right|_0
  +
  \left. \frac{d^2E^\text{II}}{d\textbf{B} d\textbf{m}^\text{X}} \right|_0
  +
  \left. \frac{d^2E^\text{(int)}}{d\textbf{B} d\textbf{m}^\text{X}} \right|_0.
  \label{eq:shielding_partitioning}
\end{equation}
The last term involves the derivatives of the interaction energy, which introduce the embedding potential and the embedding kernel contributions to the quantities calculated in the linear response equations and to the expectation value part of the shielding tensor.\cite{morbaspgpccp19}

An alternative presentation of the NMR shielding constant in subsystem-based context stems from the partitioning of the density of a supermolecule (Eq.~\ref{eq:rho_partitioning}) and 
from describing each subsystem by its own set of externally-orthogonal orbitals, which implies that
\begin{equation}
  \sigma_\text{super}(\text{X}) = \sigma^\text{I}(\text{X}) + \sigma^\text{II}(\text{X}).
  \label{eq:shielding_partitioning2}
\end{equation}
This formulation emphasizes that the shielding constant of nucleus X in the total system, $\sigma_\text{super}(\text{X})$, arises as a response of both densities (composing the total density) to the magnetic perturbations.
In particular, 
$\sigma^\text{I}(\text{X})$ -- the shielding of X in 
the active subsystem 
which contains this nucleus  -- involves the contributions from the derivatives of $E^\text{I}[\rho^\text{I}]$ as well as the functional derivatives of $E^\text{(int)}[\rho^\text{I}, \rho^\text{II}]$ with respect to the density of the active subsystem.
The shielding of X originating 
from the response in the environment (which does not contain X)
-- $\sigma^\text{II}(\text{X})$ -- can be interpreted and estimated by using an idea similar to NICS;\cite{pvrscmadhjnjrvejhjacs118}
the value of $\sigma^\text{II}(\text{X})$ 
represents 
the effect of the magnetically-induced current density in 
the environment 
at the position of X in 
the active subsystem.
This becomes more intuitive once we recall that
the NMR shielding tensor of the nucleus X can be considered as a measure of 
the magnetically-induced current probed by the nuclear magnetic dipole moment 
of that nucleus. Therefore, despite the local nature of the nuclear magnetic 
dipole moment operator which would suggest the unimportance 
of $\sigma^\text{II}(\text{X})$, the presence of neighboring 
subsystems exhibiting significant induced currents (e.g. 
aromatic\cite{bulo-jpca-112-2640-2008} or anti-aromatic) or subsystems 
that strongly overlap with the active subsystem, may lead to a non-negligible contribution to the value of the shielding constant of X.
This motivates a yet another presentation of $\sigma_\text{super}(\text{X})$ of Eq.~\ref{eq:shielding_partitioning2} as
\begin{eqnarray}
  \sigma_\text{super}(\text{X}) &= ~\sigma^\text{I}_\text{isol}(\text{X}) + \sigma^\text{I,(I)}_\text{int}(\text{X}) + \sigma^\text{I,(II)}_\text{int}(\text{X}) \label{eq:shielding_partitioning3_a} \\
                                &+ ~\sigma^\text{II}_\text{isol}(\text{X}) + \sigma^\text{II,(II)}_\text{int}(\text{X}) + \sigma^\text{II,(I)}_\text{int}(\text{X}) \label{eq:shielding_partitioning3_b}
\end{eqnarray}
in which the shielding of the isolated subsystem, $\sigma^\text{I}_\text{isol}(\text{X})$, is augmented with several contributions due to the presence of neighboring subsystems (K,L $\in$ \{I, II\}), 
\begin{eqnarray}
  \boldsymbol\sigma^\text{K,(L)}_\text{int}(\text{X}) 
  &= 
  \iint
  \left. \frac{\delta^2E^\text{(int)}}{\delta \rho_{k}^\text{K}(\textbf{r}_1) \delta \rho_{k'}^\text{L}(\textbf{r}_2)}\right|_0
  \frac{\partial \rho_{k}^\text{K}(\textbf{r}_1)}{\partial \textbf{m}^\text{X}}
  \frac{\partial \rho_{k'}^\text{L}(\textbf{r}_2)}{\partial \textbf{B}}d\textbf{r}_1 d\textbf{r}_2. \label{eq:shielding_partitioning3_int1_a} \\[1em]
  &+
  \delta_\text{KL}\int
  \left. \frac{\delta E^\text{(int)}}{\delta \rho_{k}^\text{K}\text(\textbf{r})}\right|_0
  \frac{\partial^2 \rho_{k}^\text{K}(\textbf{r})}{\partial \textbf{B} \partial \textbf{m}^\text{X}}d\textbf{r}
\label{eq:shielding_partitioning3_int1_b}
\end{eqnarray}
including two types of contributions arising from the interaction energy term: one involving its functional derivatives with respect to the general density component of the same subsystem (K=L), and the other corresponding to its \emph{mixed} functional derivatives with respect to the active and the frozen general density components (K$\neq$L).

What we aim at in this work is to 
analyse 
the performance of FDE for the description of solvent effects on $\sigma\text{(X)}$.
In practice, we will test two approximations for that purpose:
\begin{itemize}
\item ``fde0'', 
which assumes that 
in all calculations
the density of the environment 
is fixed and frozen (as an isolated subsystem II) 
and 

\item ``fde(N)'', a freeze-and-thaw 
procedure, which allows to iteratively relax the densities of both subsystems (in N cycles, in principle 
until self-consistency)  prior to property calculations.

\end{itemize}

In each case, the embedding potential is added to the KS potential in the KSCED equations.
Applying the freeze-and-thaw  procedure in the FDE scheme one may 
in principle 
reproduce exactly the corresponding 
KS-DFT supermolecular results
(which for us constitute the proper benchmark values),
provided that the same approximate exchange-correlation potential is used in both calculations and that the exact embedding potential is employed in the former.
In its practical realization, few caveats should be mentioned: (i) two non-classical contributions 
to the interaction energy -- due to the non-additive exchange-correlation energy 
and the non-additive kinetic energy 
-- are not known and have to be approximated with the density functionals and (ii) all contributions arising from the derivatives of the interaction energy should be considered -- including the \emph{mixed} derivatives with respect to active and frozen densities, responsible for the coupling not only of the ground-state densities, but also of their responses in consecutive property calculations.

\subsection{Computational aspects}\label{computational-setup}

We make use of the equilibrium structures of the transition metal complexes with the explicit
first solvent shell which have been optimized~\cite{aamrpccp22} at the 
DFT level using the PBE0 functional.~\cite{dft:pbe0}
For atoms belonging to the complexes the Def2-QZVP~\cite{fwrapccp7} basis set
and for solvent molecules the double-$\zeta$ basis set
were  used in the optimization.
The effect of the rest of the solvent has been modeled by the implicit solvent
COSMO model.
The resulting equilibrium transition metal complex geometries are in a good agreement with experimental data (for more details see Ref.~\cite{aamrpccp22}).
The complexes have been modeled including either six (C$_6$H$_6$, Cd(CH$_3$)$_2$, Hg(CH$_3$)$_2$) or twelve (TiCl$_{4}$, H$_2$O,  CH$_3$CN) solvent molecules.

The calculations of the NMR shielding constants at the spin-orbit zeroth-order regular approximation (SO-ZORA level)\cite{vanlenthe-jcp-99-4597-1993,vanlenthe-jcp-101-9783-1994} 
were performed using the ADF software (version 2019.301)\cite{ADF2001,adf2019} with the Slater-type TZP basis on all atoms, augmented by the diffuse QZ4P functions added for the fit procedure. These calculations were 
facilitated by the PyADF program package.\cite{jacob-jcc-32-2328-2011}

Four-component relativistic calculations were performed within the development version of the DIRAC software (version 7aaf401).\cite{doi:10.1063/5.0004844,DIRAC19} The Dirac--Coulomb (DC) Hamiltonian and uncontracted ANO-RCC~\cite{borrlpamvvpowjpca108} basis set for the heavy metal combined with 
uncontracted cc-pVDZ~\cite{thdjcp90} basis for all the other atoms in both subsystems were used. 
This combination of basis sets was  previously recommended
for these systems\cite{aamrpccp22}.
The simple magnetic balance scheme was used to generate the exponents for the small-component basis.\cite{olejniczak-jcp-136-014108-2012} 
The (SS$|$SS)-type two-electron integrals were approximated by a Coulombic correction.\cite{lvtcan98}
DIRAC calculations were restricted to complexes in aqueous environment.

B3LYP\cite{dft:b3lyp}
was chosen as the exchange--correlation functional for all calculations at both SO-ZORA and DC levels
(we have verified that at the 
SO-ZORA level the PBE0 functional gives for the complexes in vacuo qualitatively similar results,
see the supplementary information).
The choice of B3LYP is motivated by its good performance on the same
molecular complexes, discussed previously.\cite{aamrpccp22} 
In addition, we have used  
BLYP\cite{dft:becke88,dft:lyp1}
functional for the non-additive exchange-correlation energy potential and PW91k\cite{lembarki-pra-50-5328-1994} functional for the non-additive kinetic energy potential (therefore FDE calculations involved B3LYP, BLYP and PW91k functionals). 
The selection of BLYP was dictated by its availability in both programs, while the choice of PW91k is based on its good performance for 
transition metal 
complexes.\cite{gotz-jctc-5-3161-2009}
Freeze-and-thaw calculations were restricted to 6 iterations, which in 
every case 
was sufficient to converge the electronic energy of the transition metal complex. 
In FDE calculations and in calculations on isolated subsystems, the molecular orbitals were expanded in monomer basis functions.

In the calculations of $\sigma_\text{isol}$, $\sigma_\text{fde0}$ and $\sigma_\text{fde(N)}$, 
aimed at approximating the $\sigma_\text{super}$ value, the contributions from the isolated frozen subsystem (the first term of Eq.~\ref{eq:shielding_partitioning3_b}) were either neglected (in DC) or estimated by considering the effect of the magnetically-induced current density in the frozen subsystems in a NICS-like manner (in SO-ZORA). The coupling terms ($\sigma_\text{int}^\text{I,(II)}$ in Eq.~\ref{eq:shielding_partitioning3_a} and $\sigma_\text{int}^\text{II,(I)}$ in Eq.~\ref{eq:shielding_partitioning3_b}) were neglected in all calculations.

All calculations employed London atomic orbitals (LAOs, otherwise known as gauge-including atomic orbitals - GIAOs),~\cite{fljpr8,kwjfhppjacs112} which 
eliminate 
gauge-origin dependence of the results obtained 
otherwise with incomplete basis sets. 
Nuclear charges were modeled by Gaussian functions. 

The  integration grid generated in ADF ("very good" quality\cite{neugebauer-jpca-109-7805-2005}) 
taking into account all atoms in the total systems was used in all calculations (ADF and DIRAC), except for the $\sigma_\text{vac}$ and the $\sigma_\text{isol}$ values which were 
computed
for isolated subsystems 
on their respective grids.

\section{Results and discussion}\label{results}

\subsection{Absolute shielding constants}
\label{subsec:absolute}

\begin{table*}[!h]
\caption{Absolute 
heavy atom
shielding constants, $\sigma$(X), calculated with the 
SO-ZORA Hamiltonian. The descriptions of computational details and of the symbols used in the table are given in the text$^a$. 
All values in ppm.}
\label{tab:adf_absolute}
\centering
 \setlength{\tabcolsep}{2pt}
\begin{tabular}{ l   l | r   r   r   r   r   r   r } 
\hline
    & Solvent 
    & $\sigma_\text{vac}$ 
    & $\sigma^\text{I}_\text{isol}$ 
    & $\sigma^\text{II}_\text{isol}$  
    & $\sigma^\text{I}_\text{fde0}$ 
    & $\sigma^\text{I}_\text{fde(N)}$ 
    & $\sigma^\text{II}_\text{fde}$ 
    & $\sigma_\text{super}$ \\
  \hline
TiCl$_4$      &   TiCl$_{4}$            & -821.8  &-799.6  &-0.5 &-800.1    &-799.9  &-0.6 &-817.5    \\  
VOCl$_3$      &  C$_6$H$_6$             & -1860.1 &-1857.3 &3.4  &-1859.1   &-1860.2 &3.4  &-1885.4   \\
CrO$_4^{2-}$    &   H$_2$O              & -2725.8 &-2602.8 &-0.1 &-2621.8   &-2649.2 &-0.3 &-2662.0   \\
MnO$_4^{-}$     &   H$_2$O              & -3934.9 &-3899.4 &0.0  & -3905.8  &-3919.7 &-0.2 &-3942.2   \\
Fe(CO)$_5$      &  C$_6$H$_6$           & -2331.0 &-2263.6 &2.8  &-2265.6   &-2266.4 &2.8  &-2283.9   \\
Co(CN)$_6^{3-}$   &   H$_2$O            & -5774.5 &-4676.2 &0.1  &-4646.9   &-4650.0 &0.0  &-4677.1   \\
Ni(CO)$_4$      &  C$_6$H$_6$           & -1629.9 &-1639.2 &2.8  &-1647.8   &-1651.3 &2.8  &-1657.4   \\
Cu(CH$_3$CN)$_4^{+}$  &   CH$_3$CN      & 477.4   &454.1   &0.1  &432.5     &429.5   &0.1  &423.9     \\
Zn(H$_2$O)$_6^{2+}$   &   H$_2$O        & 1914.1  &1884.4  &-0.1 &1883.8    &1884.0  &-0.1 &1871.0    \\
&&&&&&&&\\
NbCl$_6^{-}$    &   CH$_3$CN            & -317.9  &-302.0  &0.1  &-314.7    &-317.8  &0.0  &-337.5    \\
MoO$_4^{2-}$    &   H$_2$O              & -559.2  &-489.7  &-0.1 &-501.7    &-521.2  &-0.2 &-542.9    \\
TcO$_4^{-}$     &   H$_2$O              & -1419.1 &-1396.4 &-0.1 &-1390.2   &-1395.2 &-0.2 &-1392.1   \\
Ru(CN)$_6^{4-}$   &   H$_2$O            & -811.5  &16.7    &0.1  & 20.3     &22.7    &-0.1 &-23.7     \\
PdCl$_6^{2-}$     &   H$_2$O            & -1580.6 &-1377.7 &0.0  &-1342.0   &-1310.1 &-0.1 &-1309.6   \\
Ag(H$_2$O)$_4^{+}$  &   H$_2$O          & 4315.0  &4139.9  &-0.2 &4140.7    &4141.5  &-0.1 &4104.8    \\
Cd(CH$_3$)$_2$    &  Cd(CH$_3$)$_2$     & 3456.2  &3463.2  &0.0  &3478.9    &3503.0  &0.0  &3511.1    \\
&&&&&&&&\\
TaCl$_6^{-}$    &   CH$_3$CN            & 3410.5  &3422.1  &0.1  &3403.7    &3399.0  &0.0  &3441.2    \\
WO$_4^{2-}$     &   H$_2$O              & 3002.0  &3087.2  &0.0  &3054.4    &3027.1  &-0.2 &2927.6    \\
ReO$_4^{-}$     &   H$_2$O              & 1892.7  &1914.3  &-0.1 &1904.2    &1895.4  &-0.2 &1839.8    \\
OsO$_4$       &   CCl$_4$               & 1218.4  &1351.1  &-0.5 &1354.0    &1356.0  &-0.6 &1339.6    \\
PtCl$_6^{2-}$     &   H$_2$O            & 1885.3  &2177.6  &0.0  &2256.4    &2313.3  &-0.1 &2383.7    \\
Hg(CH$_3$)$_2$    &  Hg(CH$_3$)$_2$     & 8882.6  &8914.5  &-0.1 &8949.8    &8978.1  &-0.1 &8922.7    \\
  \hline
\end{tabular}

$^a$The values 
denoted as $\sigma^\text{II}_\text{fde}$ represent 
$\sigma^\text{II}_\text{fde0}$ and $\sigma^\text{II}_\text{fde(N)}$,
which  are the same within the given accuracy, 
Their contributions to the total shielding constants are practically negligible, thus we do not 
discuss them in the text.
\end{table*}

\begin{table*}[!h]
\caption{Absolute 
heavy atom
shielding constants, $\sigma$(X),  
calculated with the DC Hamiltonian
for H$_2$O complexes. The descriptions of computational details and of the symbols used in the table are given in the text. All values in ppm.}
\label{tab:dirac_absolute}
\centering
 \setlength{\tabcolsep}{2pt}
\begin{tabular}{ l    | r   r   r   r   r } 
\hline
    & $\sigma_\text{vac}$ 
    & $\sigma^\text{I}_\text{isol}$ 
    & $\sigma^\text{I}_\text{fde0}$ 
    & $\sigma^\text{I}_\text{fde(N)}$ 
    & $\sigma_\text{super}$ \\
  \hline
CrO$_4^{2-}$                            & -2944.3&-2813.9&-2796.4&-2816.3&-2863.7  \\
MnO$_4^{-}$                             & -4259.9&-4221.8&-4225.8&-4236.5&-4279.8  \\
Co(CN)$_6^{3-}$                         & -6125.7&-4967.6&-4963.6&-4966.1&-4993.8  \\
Zn(H$_2$O)$_6^{2+}$                     & 1985.4&1947.7&1946.9&1946.9&1939.1       \\
&&&&&\\
MoO$_4^{2-}$                            & -403.3&-335.6&-315.8&-324.5&-383.7       \\
TcO$_4^{-}$                             & -1264.3&-1241.6&-1228.4&-1230.4&-1254.5  \\
Ru(CN)$_6^{4-}$                         & -762.9&74.8&205.0& 220.3&     228.3            \\
PdCl$_6^{2-}$                           & -1768.4&-1550.4&-1532.9&-1501.8&-1477.3  \\
Ag(H$_2$O)$_4^{+}$                      & 4589.1&4408.0&4408.3&4407.4&4377.6       \\
&&&&&\\
WO$_4^{2-}$                             & 4468.9&4547.6&4567.5&4557.9&4436.9    \\
ReO$_4^{-}$                             & 3414.6&3435.8&3446.9&3442.5&3390.2    \\
PtCl$_6^{2-}$                           & 2557.5&2884.9&3066.9&3125.7& 3376.8   \\
  \hline
\end{tabular}
\end{table*}

The calculated values of the absolute shielding constants of heavy atoms in 
the set of studied systems 
are summarized in Table~\ref{tab:adf_absolute} (SO-ZORA) and Table~\ref{tab:dirac_absolute} (DC).

The transition metal nuclei of the 
3d and 4d elements 
of the periodic table 
in negatively-charged complexes adopt negative $\sigma_\text{vac}$ values, while those of the 5d elements
in all considered complexes display large positive shielding constants.
The transition metal nuclei in positively-charged complexes 
have
positive shielding constants (Cu, Zn and Ag complexes), while those in neutral complexes manifest mixed $\sigma_\text{vac}$ values, with a tendency to adopt negative values for lighter metals (Ti, V, Fe and Ni) and positive values for heavier ones (Cd, Os, Hg).

In this respect, both SO-ZORA and DC calculations lead to the same qualitative conclusions. Yet, a closer look at the results obtained with these two Hamiltonians paints a more nuanced picture.
For instance, for the complexes \emph{in vacuo}
- considering the DC values as the most accurate - 
the error of 
the approximate SO-ZORA approach,
calculated as $e_\text{vac} = 100\times(\sigma_\text{vac}\text{(SO-ZORA)} - \sigma_\text{vac}\text{(DC)})/\sigma_\text{vac}\text{(DC)}$,
can be as large as 
-33\% in W and -45\% in Re complexes ($e_\text{vac}$ values for complexes in water are demonstrated in Table~\ref{tab:esi-zora-dc}).
There is no obvious  correlation of $e_\text{vac}$ with the charge of the complex or the atomic number of the nucleus of interest 
for the 
3d and 4d
elements. 
However, in all complexes involving 
5d transition metals,
the SO-ZORA values significantly underestimate the DC reference.

The signs of $\sigma_\text{super}$ of transition metals in solvents follow 
similar trends as the signs of $\sigma_\text{vac}$
for complexes of different charge.
The comparison of $\sigma_\text{super}$ of selected transition metals obtained with SO-ZORA and DC Hamiltonians does not lend itself to systematic trends as well. The differences between the results for these two Hamiltonians 
are similar to the ones for transition metal complexes 
\emph{in vacuo} ($e_\text{super}$ in Table~\ref{tab:esi-zora-dc}).

\begin{table*}[!h]
\caption{The percentage values of errors of the SO-ZORA results for heavy atoms with respect to the DC values, calculated as $e = 100\times(\sigma\text{(SO-ZORA)} - \sigma\text{(DC)})/\sigma\text{(DC)}$, for supermolecules ($e_\text{super}$), complexes in vacuo ($e_\text{vac}$), and for complexes with estimated mechanical solvent effects ($e_\text{isol}$),  solvent effects estimated at fde0 ($e_\text{fde0}$) and fde(N) ($e_\text{fde(N)}$) levels.$^a$ }
\label{tab:esi-zora-dc}
\centering
 \setlength{\tabcolsep}{2pt}
\begin{tabular}{ l | r   r   r   r   r } 
\hline
    & $e_\text{super}$
    & $e_\text{vac}$ 
    & $e_\text{isol}$ 
    & $e_\text{fde0}$ 
    & $e_\text{fde(N)}$  \\
  \hline
CrO$_4^{2-}$           & -7.0  & -7.4 & -7.5 & -6.2 & -5.9 \\ 
MnO$_4^{-}$            & -7.9  & -7.6 & -7.6 & -7.6 & -7.5 \\ 
Co(CN)$_6^{3-}$        & -6.3  & -5.7 & -5.9 & -6.4 & -6.4 \\ 
Zn(H$_2$O)$_6^{2+}$    & -3.5  & -3.6 & -3.2 & -3.2 & -3.2 \\ 
&&&&&\\
MoO$_4^{2-}$           & 41.5 & 38.7  & 45.9 & 58.9 & 60.6 \\ 
TcO$_4^{-}$            & 11.0 & 12.2  & 12.5 & 13.2 & 13.4 \\ 
Ru(CN)$_6^{4-}$        &-110.4&  6.4  &-77.7 &-90.1 & -89.7 \\ 
PdCl$_6^{2-}$          &-11.4 &-10.6  &-11.1 &-12.5 & -12.8 \\ 
Ag(H$_2$O)$_4^{+}$     & -6.2 & -6.0  & -6.1 & -6.1 & -6.0 \\ 
&&&&&\\
WO$_4^{2-}$            &-34.0 &-32.8  &-32.1 &-33.1 & -33.6 \\ 
ReO$_4^{-}$            &-45.7 &-44.6  &-44.3 &-44.8 & -44.9 \\ 
PtCl$_6^{2-}$          &-29.4 &-26.3  &-24.5 &-26.4 & -26.0 \\ 
  \hline
\end{tabular}

$^a$ Results are available for complexes in water.
\end{table*}

\par\null

\subsection{Solvent effects}
\label{subsec:solveff}

For the 
discussion of the computed solvent effects on the shielding 
we introduce the solvent shifts defined as 
\begin{equation}
  \delta =  \sigma - \sigma_\text{vac}
  \label{eq:delta}
\end{equation}
where $\sigma_\text{vac}$ corresponds to the shielding constant in the \emph{in vacuo} complex, and $\sigma$ - to the corresponding shielding constant in the solvent calculated with one of the approximations explored in this work. In particular,
\begin{itemize}
\item selecting $\sigma = \sigma_\text{isol}$ allows to estimate $\delta_\text{isol}$, which represents the solvent shift driven by relaxation of 
the geometry of the active complex to the geometry it acquires
in a supermolecule,
\item selecting $\sigma = \sigma_\text{fde0}$ results in $\delta_\text{fde0}$ representing the solvent shift due to the geometry relaxation and to the electronic effects captured by the fde0 approximation,
\item selecting $\sigma = \sigma_\text{fde(N)}$ allows to evaluate $\delta_\text{fde(N)}$ embodying the solvent shift due to the geometry relaxation and to the electronic effects captured by the fde(N) approximation,
\item finally $\delta_\text{super}$, calculated with $\sigma = \sigma_\text{super}$, represents the supermolecular solvent shift.
\end{itemize}
We shall assess the solvent shift approximations assuming $\delta_\text{super}$ to be the exact reference value;
the systematic error of $\delta_\text{super}$ cannot be determined, because it is impossible to perform a direct
measurement of absolute solvent shifts for most of the studied complexes.

\begin{table*}[!h]
\caption{Solvent shifts on 
heavy atom
$\sigma$(X) using SO-ZORA Hamiltonian. The descriptions of computational details and of the symbols used in the table are given in the text. All values in ppm.}
\label{tab:zora-deltas}
\centering
 \setlength{\tabcolsep}{2pt}
\begin{tabular}{ l   l |  r   r   r   r  } 
\hline
    & Solvent
    & $\delta_\text{isol}$ 
    & $\delta_\text{fde0}$ 
    & $\delta_\text{fde(N)}$ 
    & $\delta_\text{super}$\\
  \hline
TiCl$_4$      &   TiCl$_{4}$        & 22.2&21.7&21.9&4.3                                 \\  
VOCl$_3$      &  C$_6$H$_6$         & 2.8&1.0&-0.1&-25.3                                 \\
CrO$_4^{2-}$    &   H$_2$O          & 123.0&104.0&76.7&63.9                              \\
MnO$_4^{-}$     &   H$_2$O          & 35.5&29.1&15.2&-7.2                                \\
Fe(CO)$_5$      &  C$_6$H$_6$       & 67.4&65.4&64.6&47.1                                \\
Co(CN)$_6^{3-}$   &   H$_2$O        & 1098.3&1127.6&1124.5&1097.4                              \\
Ni(CO)$_4$      &  C$_6$H$_6$       & -9.4&-18.0&-21.5&-27.5                             \\
Cu(CH$_3$CN)$_4^{+}$  &   CH$_3$CN  & -23.3&-44.9&-47.9&-53.5                            \\
Zn(H$_2$O)$_6^{2+}$   &   H$_2$O    & -29.7&-30.3&-30.1&-43.1                            \\
&&&&&\\
NbCl$_6^{-}$    &   CH$_3$CN        & 15.9&3.2&0.2&-19.5                                       \\
MoO$_4^{2-}$    &   H$_2$O          & 69.5&57.5&38.0&16.3                                      \\
TcO$_4^{-}$     &   H$_2$O          & 22.7&28.9&23.9&27.0                                      \\
Ru(CN)$_6^{4-}$   &   H$_2$O        & 828.1&831.8&834.2&787.8                                  \\
PdCl$_6^{2-}$     &   H$_2$O        & 202.9&238.6&270.5&271.0                                  \\
Ag(H$_2$O)$_4^{+}$  &   H$_2$O      & -175.0&-174.3&-173.4&-210.2                              \\
Cd(CH$_3$)$_2$    &  Cd(CH$_3$)$_2$ & 7.0&22.7&46.8&54.9                                       \\
&&&&&\\
TaCl$_6^{-}$    &   CH$_3$CN        & 11.6&-6.8&-11.6&30.6                       \\
WO$_4^{2-}$     &   H$_2$O          & 85.2&52.4&25.1&-74.4                       \\
ReO$_4^{-}$     &   H$_2$O          & 21.6&11.6&2.7&-52.8                        \\
OsO$_4$       &   CCl$_4$           & 132.7&135.6&137.6&121.2                    \\
PtCl$_6^{2-}$     &   H$_2$O        & 292.3&371.2&428.1&498.5                    \\
Hg(CH$_3$)$_2$    &  Hg(CH$_3$)$_2$ & 32.0&67.2&95.5&40.1                  \\
  \hline
\end{tabular}
\end{table*}

\begin{table*}[!h]
\caption{Solvent shifts on 
heavy atom
$\sigma$(X) for H$_2$O
complexes calculated using DC Hamiltonian. The descriptions of computational details and of the symbols used in the table are given in the text. All values in ppm.}
\label{tab:dc-deltas}
\centering
 \setlength{\tabcolsep}{2pt}
\begin{tabular}{ l  |  r   r   r   r } 
\hline
    & $\delta_\text{isol}$ 
    & $\delta_\text{fde0}$ 
    & $\delta_\text{fde(N)}$ 
    & $\delta_\text{super}$\\
  \hline
CrO$_4^{2-}$       &130.4&147.8&128.0&80.6       \\
MnO$_4^{-}$        &38.0&34.1&23.4&-20.0           \\
Co(CN)$_6^{3-}$    &1158.1&1162.1&1159.6&1131.9  \\
Zn(H$_2$O)$_6^{2+}$&-37.7&-38.6&-38.5&-46.3    \\
MoO$_4^{2-}$       &67.7&87.5&78.8&19.6                  \\
TcO$_4^{-}$        &22.7&35.9&33.9&9.8                   \\
Ru(CN)$_6^{4-}$    &837.7&967.9&983.2&  991.2         \\
PdCl$_6^{2-}$      &218.0&235.5&266.6&291.1          \\
Ag(H$_2$O)$_4^{+}$ &-181.0&-180.8&-181.7&-211.4  \\
WO$_4^{2-}$        &78.7&98.6&89.0&-31.9  \\
ReO$_4^{-}$        &21.2&32.3&27.9&-24.3   \\
PtCl$_6^{2-}$      &327.5&509.4&568.2&819.3  \\
  \hline
\end{tabular}
\end{table*}

\begin{figure}[h!]
\begin{center}
    \includegraphics[scale=0.3]{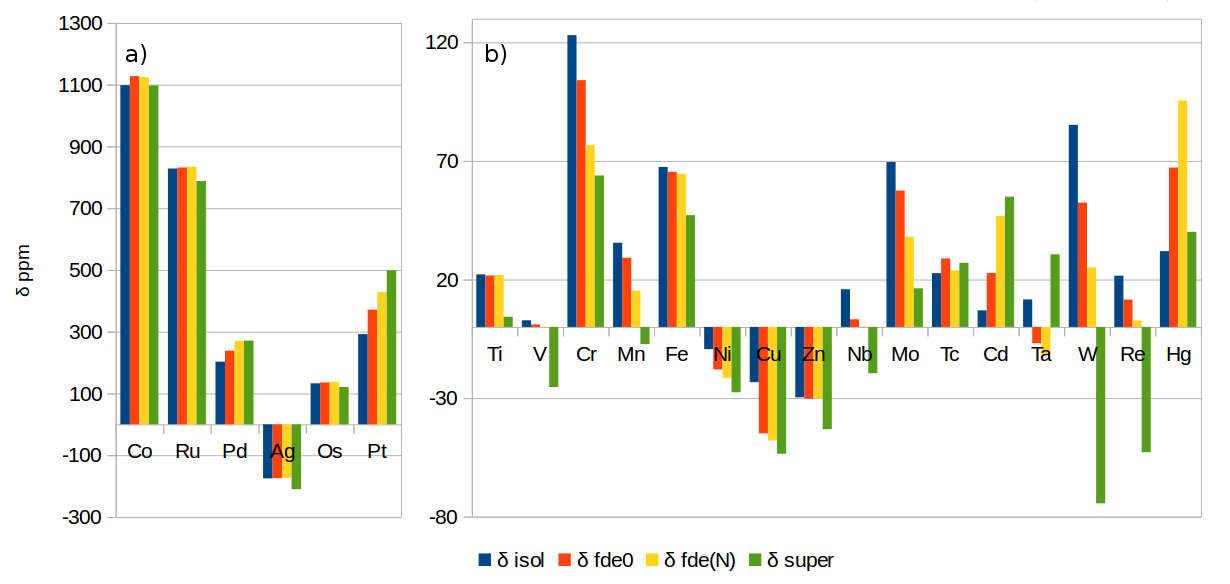}
\caption{SO-ZORA solvent shifts 
$\delta_\text{isol}$, $\delta_\text{fde0}$, $\delta_\text{fde(N)}$ and $\delta_\text{super}$  
calculated 
for a) systems with $|\delta_\text{super}|$  larger than 150 ppm. b) systems with  $|\delta_\text{super}|$ smaller than 150 ppm.
}
\label{fig:deltas-zora-merge}
\end{center}
\end{figure}

\paragraph{Comparison of SO-ZORA and DC solvent shifts}

Due to the expected cancellation of errors
a better agreement between SO-ZORA and DC results is typically obtained for the differences of the absolute shielding constants - for instance for the chemical shifts.\cite{autschbach-mp-111-2544-2013} 
This is also observed to some extent for the solvent shifts calculated in this work. As presented in Tables~\ref{tab:zora-deltas} and \ref{tab:dc-deltas}, the overall trends for accurate solvent shifts ($\delta_\text{super}$) and for solvent shifts 
determined
by the FDE scheme 
in fde0 or in fde(N) variants ($\delta_\text{fde0}$ and $\delta_\text{fde(N)}$, respectively) at the SO-ZORA and DC levels agree rather well. 
This agreement is not as close as 
could 
be
expected based on experience with chemical shifts,\cite{autschbach-mp-111-2544-2013}
however, non-negligible differences in the description of solvent effects offered by these two Hamiltonians were already observed in smaller systems with heavy nuclei.\cite{morbaspgpccp19}

In particular,  accurate solvent shifts predicted with these two approaches 
differ for the studied anionic four-coordinated transition metal 
oxides in water. In case 
of heavy 5d  transition metals, the SO-ZORA values of $\delta_\text{super}$ are significantly 
larger (in absolute terms) than the DC $\delta_\text{super}$ estimates 
(over 2 times larger in W and Re complexes), while it is the opposite for 3d elements Cr and Mn, 
in which the absolute values of $\delta_\text{super}$ predicted by the SO-ZORA Hamiltonian are smaller than the
more accurate 
four-component values. Among the intermediate 4d elements, the accurate solvent shifts of Mo in MoO$_4^{2-}$ + 12 H$_2$O are
substantially larger when predicted with the SO-ZORA Hamiltonian than when the DC framework is used, while the opposite trend is observed for Ru in Ru(CN)$_6^{4-}$ + \mbox{12 H$_2$O}.
Other complexes in water do not manifest significant discrepancies of $\delta_\text{super}$ values obtained with these two Hamiltonians.

In order to better analyze the FDE results, the SO-ZORA solvent shifts collected in  Table~\ref{tab:zora-deltas} are also presented separately for systems
with the solvent shifts larger (Figure~\ref{fig:deltas-zora-merge}a) and smaller (Figure~\ref{fig:deltas-zora-merge}b) 
than 150 ppm.   
Taking into account the nature of 'isol', 'fde0' and 'fde(N)' approximations, monotonous convergence of $\delta_\text{isol}$, $\delta_\text{fde0}$, $\delta_\text{fde(N)}$ values towards the reference $\delta_\text{super}$ solvent shift is expected.
For the systems with largest solvent shifts
%{\color{red}
%, which involve electron-rich ligands
%}
(Figure~\ref{fig:deltas-zora-merge}a), this series of approximations either leads to a systematic improvement of computed solvent shifts (e.g. for  PdCl$_6^{2-}$ and PtCl$_6^{2-}$ complexes in 
water), or results in equally good estimations of these shifts.
However, when we consider small solvent shifts we observe a few problematic cases in our test set. For WO$_4^{2-}$ and ReO$_4^{2-}$ complexes in water 
we find negative reference solvent shifts $\delta_{super}$, whereas the  
$\delta_\text{isol}$, $\delta_\text{fde0}$, $\delta_\text{fde(N)}$ series of approximations predicts positive solvent shifts, 
but
the trend in the series is correct. 
A different pattern is observed for Ta(Cl)$_6^-$ and Hg(CH$_3$)$_2$ systems only, where the sequence of $\delta_\text{isol}$, $\delta_\text{fde0}$ and $\delta_\text{fde(N)}$ does not converge to reference supermolecular shift $\delta_\text{super}$.

SO-ZORA $\delta_\text{fde(N)}$ predicts the sign of the solvent shift correctly for most complexes in our test set, 
the most noticeable failures  are the W and Re complexes. We recall here that the solvent shift is computed as the difference of two larger numbers (in case of W and Re thousands of ppm),
and the performance of the applied approximations for the  isolated complex and for the solvated complex may differ.
Nevertheless, SO-ZORA $\delta_\text{fde(N)}$  approximation predicts the solvent shifts within 
20-30 ppm accuracy in most cases. 
\color{black}

The DC solvent shifts are presented in Table~\ref{tab:dc-deltas}. The monotonous convergence of $\delta_\text{isol}$, $\delta_\text{fde0}$, $\delta_\text{fde(N)}$ values towards the reference $\delta_\text{super}$ values is observed only for MnO$_4^{2-}$, Ru(CN)$_6^{4-}$, Pd(Cl)$_6^{2-}$ and Pt(Cl)$_6^{2-}$ complexes. The breakdown of the monotonous behavior in other systems is due to the fact that $\delta_\text{fde0}$ values significantly overestimate the reference shifts, yet the correct trend is restored by the fde(N) approach.

For water solvated complexes, DC fde(N) predicts the signs of the solvent shifts consistent with SO-ZORA fde(N)
approximation. However, the discrepancy between fde(N) and the corresponding reference supermolecular 
result,
$|\delta_\text{fde(N)}$ -  $\delta_\text{super}|$, is about two times larger for the DC method than for the SO-ZORA method.

\begin{figure*}[!h]
\centering
    \includegraphics[scale=0.30]{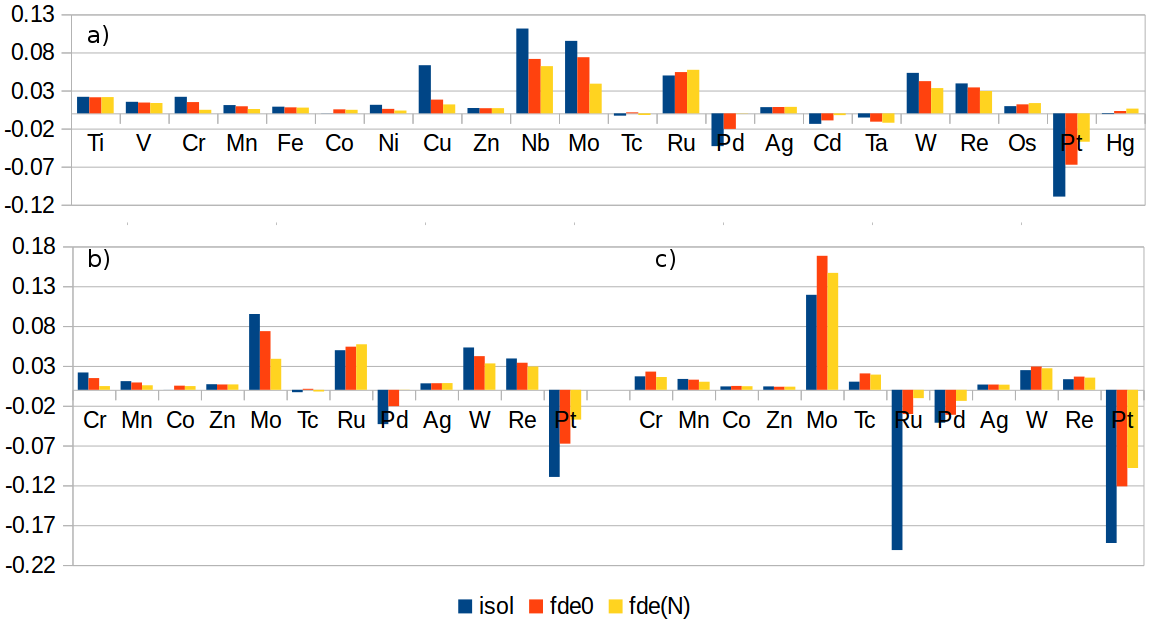}
\caption{Relative discrepancies between reference $\delta_\text{super}$ and $\delta_\text{isol}$, $\delta_\text{fde0}$ and $\delta_\text{fde(N)}$ solvent shifts, calculated using Eq.~\ref{disc_f}. a) SO-ZORA all systems, b) SO-ZORA water solvated systems, c) DC water solvated systems}
\label{fig:discrep-multiplot}
\end{figure*}

Absolute NMR shielding constants are important for the derivation 
of nuclear magnetic dipole moments~\cite{aamrpccp22} from NMR experiments. Therefore, it
is instructive to estimate the relative error in absolute shielding constants introduced by
different approaches to solvent shifts. In order to analyze the performance
of FDE in this context, we define a relative discrepancy $d_r$ measured
with respect to the solvent-independent absolute value of the shielding in vacuum $|\sigma_\text{vac}|$,
\begin{equation}
d_r = \frac{\delta-\delta_\text{super}}{|\sigma_\text{vac}|} = \frac{\sigma-\sigma_\text{super}}{|\sigma_\text{vac}|},
\label{disc_f}
\end{equation}
where $\sigma$ and $\delta$ refer to one of the isol, fde0 and fde(N) models.
The relative discrepancies of isol, fde0 and fde(N) approximations for SO-ZORA method 
are shown in Figure~\ref{fig:discrep-multiplot}a.
We note that fairly large $d_r$ discrepancies for Cu, Nb and Mo complexes reflect the smallest values of the reference  ${|\sigma_\text{vac}|}$ for these complexes.
In Figure~\ref{fig:discrep-multiplot}b,c, relative discrepancies for SO-ZORA and DC method are compared for the subset of
water solvated systems. 
The calculated average (avg) discrepancy and the standard deviation 
(std) are shown 
in Table~\ref{tab:disc_tab}
for the whole set and separately for systems with positive and negative discrepancies, 
and the largest positive (max) and the largest negative (min) discrepancy is identified.
The statistics is calculated separately for SO-ZORA-based results for the complete set 
of complexes 
as well as for 
the SO-ZORA and for the DC results for water solvated systems.

The discrepancy statistics shows that isol, fde0 and fde(N) series of approximations provides an improving description of the 
solvent effects, toward the reference supermolecular results for all tested subsets except the positive values subset for DC method. 
However, the statistics of this positive subset for DC method is strongly affected by large discrepancies of Mo complex,
which stems from the small value of ${|\sigma_\text{vac}|}$. 
The iterative freeze-and-thaw procedure fde(N) improves noticeably the agreement 
with the reference supermolecular solvent shifts in comparison to the fde0 procedure. This is 
mainly due to the capability of fde(N) procedure to reduce the extreme discrepancies 
(see Figure~\ref{fig:discrep-multiplot} and Table~\ref{tab:disc_tab}).
Generally, all the extreme discrepancies come from the water solvated systems subset,  demonstrating that water is a difficult solvent to model.  
We  also note that the improvement within isol, fde0 and fde(N) series of approximations is not guaranteed
for every system.
Although the extreme discrepancies indicate non-negligible
errors in the fde approximations,  we find - based on the data of Ref.~\cite{aamrpccp22} - that the average and extreme discrepancies between PCM solvent model and supermolecular solvent model in the studied systems reached much larger values (as shown in  Table~\ref{tab:disc_tab}).

\begin{table*}[!h]
\caption{Discrepancy $d_r$ statistics
for heavy atoms.
}
\label{tab:disc_tab}
\centering
 \setlength{\tabcolsep}{2pt}
\begin{tabular}{ l   r  r   r  r   r   r   r    r   r    r} 
\hline
&\multicolumn{2}{c}{all $d_r$  values}  &&\multicolumn{3}{c}{positive $d_r$ values}&& \multicolumn{3}{c}{negative $d_r$ values} \\
\cline{2-3}\cline{5-7}\cline{9-11}
               &$avg$&$std$&& $avg$ & $std$&$max$&&$avg$& $std$&$min$\\

all complexes\\                                                                            
SO-ZORA isol      & 0.016  & 0.045  &&  0.033    &  0.033   & 0.111   && -0.029  &  0.042   & -0.109  \\
SO-ZORA fde0      & 0.013  & 0.030  &&  0.022    &  0.023   & 0.074   && -0.027  &  0.027   & -0.068  \\
SO-ZORA fde(N)    & 0.012  & 0.022  &&  0.018    &  0.019    & 0.062  && -0.011   &  0.016   & -0.037  \\
\\
H$_2$O complexes\\                                                                      
SO-ZORA isol      & 0.011  & 0.051  &&  0.032    &  0.032   & 0.095   &&-0.052   &  0.053   & -0.109  \\
SO-ZORA fde0      & 0.013  & 0.036  &&  0.025    &  0.025   & 0.074   &&-0.044   &  0.033   & -0.068  \\
SO-ZORA fde(N)    & 0.012  & 0.024  &&  0.021    &  0.019   & 0.057   &&-0.013   & 0.021    & -0.037  \\
DC isol      &  -0.018  & 0.091  &&  0.024    &  0.036   & 0.119   &&-0.145   & 0.090    & -0.201  \\
DC fde0        & 0.009  & 0.065  &&  0.032    &  0.052   & 0.168   &&-0.061   & 0.052   & -0.121  \\
DC fde(N)      & 0.011  & 0.054  &&  0.028    &  0.045   & 0.147   &&-0.041   & 0.050    & -0.098  \\
\\
PCM$^a$        & -0.017 & 0.108  &&  0.162    &  0.124   &  0.287  && -0.035  &  0.047   &  -0.163\\
\hline
\end{tabular}\\
$^a$ The discrepancy between PCM and supermolecular solvent shifts, calculated 
using an analogue of Eq.~\ref{disc_f}
(based on the data of Ref.~\cite{aamrpccp22})
\end{table*}

\paragraph{Contributions to solvent effects}

More insight into these solvent shifts can be gained from the analysis of different contributions to $\delta$ values. 
These contributions,
measured by the $\Delta$ values defined as 
 
\begin{align}
  \Delta_\text{isol}    &=  \sigma_\text{isol} - \sigma_\text{vac} \equiv \delta_\text{isol}   \label{eq:Delta_isol} \\
  \Delta_\text{fde0}    &=  \sigma_\text{fde0} - \sigma_\text{isol}  \label{eq:Delta_fde0} \\
  \Delta_\text{fde(N)}  &=  \sigma_\text{fde(N)} - \sigma_\text{fde0}  \label{eq:Delta_fde(N)} \\
  \Delta_\text{super}   &=  \sigma_\text{super} - \sigma_\text{fde(N)}  \label{eq:Delta_super}
\end{align}
are summarized in 
Table~\ref{tab:esi-Deltas}.
\begin{table*}[!htb]
\caption{Contributions to 
heavy atom
solvent shifts calculated from Eqs.~\ref{eq:Delta_isol}-\ref{eq:Delta_super} 
with the SO-ZORA  and DC  Hamiltonians. All values in ppm.}
\label{tab:esi-Deltas}
\centering
 \setlength{\tabcolsep}{2pt}
\begin{tabular}{ l   l | r   r   r   r | r   r   r   r } 
\hline
    & Solvent
    & \multicolumn{4}{  c  |}{SO-ZORA} 
    & \multicolumn{4}{  c  }{DC$^a$} \\
    &
    & $\Delta_\text{isol}$
    & $\Delta_\text{fde0}$
    & $\Delta_\text{fdeN}$
    & $\Delta_\text{super}$
    & $\Delta_\text{isol}$
    & $\Delta_\text{fde0}$
    & $\Delta_\text{fdeN}$
    & $\Delta_\text{super}$  \\
  \hline
TiCl$_4$              &   TiCl$_{4}$    & 22.2&-0.5&0.2&-17.6    
& & & & \\ 
VOCl$_3$              &  C$_6$H$_6$     & 2.8&-1.8&-1.1&-25.2    
& & & & \\ 
CrO$_4^{2-}$          &   H$_2$O        & 123.0&-19.0&-27.4&-12.8& 130.4&17.4&-19.9&-47.4 \\
MnO$_4^{-}$           &   H$_2$O        & 35.5&-6.4&-13.9&-22.5  & 38.0&-3.9&-10.7&-43.3 \\
Fe(CO)$_5$            &  C$_6$H$_6$     & 67.4&-2.0&-0.8&-17.5   
& & & & \\ 
Co(CN)$_6^{3-}$       &   H$_2$O        & 1098.3&29.3&-3.1&-27.1 & 1158.1&4.0&-2.5&-27.7 \\ 
Ni(CO)$_4$            &  C$_6$H$_6$     & -9.4&-8.6&-3.5&-6.1    
& & & & \\ 
Cu(CH$_3$CN)$_4^{+}$  &   CH$_3$CN      & -23.3&-21.6&-3.0&-5.6  
& & & & \\ 
Zn(H$_2$O)$_6^{2+}$   &   H$_2$O        & -29.7&-0.6&0.2&-13.0   & -37.7&-0.9&0.1&-7.8\\ 
&&&&
&&&&&\\
NbCl$_6^{-}$          &   CH$_3$CN      & 15.9&-12.7&-3.0&-19.7 & & & & \\ 
MoO$_4^{2-}$          &   H$_2$O        & 69.5&-12.0&-19.5&-21.8& 67.7&19.8&-8.7&-59.2\\ 
TcO$_4^{-}$           &   H$_2$O        & 22.7&6.2&-5.0&3.1     & 22.7&13.2&-2.0&-24.1 \\ 
Ru(CN)$_6^{4-}$       &   H$_2$O        & 828.1&3.6&2.4&-46.4   & 837.7&130.2&15.3&7.9 \\ 
PdCl$_6^{2-}$         &   H$_2$O        & 202.9&35.7&31.9&0.5   & 218.0&17.5&31.1&24.5 \\ 
Ag(H$_2$O)$_4^{+}$    &   H$_2$O        & -175.0&0.8&0.8&-36.7  & -181.0&0.2&-0.9&-29.7 \\ 
Cd(CH$_3$)$_2$        &  Cd(CH$_3$)$_2$ & 7.0&15.7&24.1&8.1     & & & & \\ 
&&&&&&&&&\\
TaCl$_6^{-}$          &   CH$_3$CN      & 11.6&-18.4&-4.7&42.2  & & & & \\ 
WO$_4^{2-}$           &   H$_2$O        & 85.2&-32.8&-27.3&-99.5& 78.7&19.9&-9.6&-120.9 \\ 
ReO$_4^{-}$           &   H$_2$O        & 21.6&-10.0&-8.8&-55.6 & 21.2&11.1&-4.4&-52.2 \\ 
OsO$_4$               &   CCl$_4$       & 132.7&2.8&2.0&-16.4   & & & & \\ 
PtCl$_6^{2-}$         &   H$_2$O        & 292.3&78.9&56.9&70.4  & 327.5&181.9&58.8&251.0 \\ 
Hg(CH$_3$)$_2$        &  Hg(CH$_3$)$_2$ & 32.0&35.2&28.3&-55.4  & & & & \\ 
  \hline
\end{tabular}

$^a$ DC results are available only for complexes in water.
\end{table*}

The analysis of the $\Delta$ values demonstrates that the solvent shifts in the studied complexes are largely driven by the change of geometry of the solute 
complexes
upon interaction with solvent molecules, with the proportionate contribution of these mechanical 
effects depending on the Hamiltonian. 
The relative importance of mechanical and electronic effects may also largely depend on the 
chosen 
geometries of these complexes, as previously observed for the MoO$_4^{2-}$ ions in water.\cite{halbert-ijqc-2020} Taking the dynamical effects into account -- for instance by means of statistical averaging 
over the range of geometries from molecular dynamics simulations -- would then be needed 
to estimate how these effects interplay in the experimental NMR shielding constants 
of these complexes in solutions.

In many studied systems the effects not covered by the FDE schemes employed here -- measured by $\Delta_\text{super}$ -- are relatively large. There are many sources of errors that can sum up to $\Delta_\text{super}$ values, such as:
(i) the 
neglect
of the coupling of responses of subsystems (``uncoupled FDE'' approach),
(ii) the errors introduced with the selection of the functionals for the non-additive exchange-correlation and the non-additive kinetic energy potentials, and
(iii) the use of monomer basis sets.

The lack of intersubsystem coupling at the response level, (i), has two direct consequences. First, it affects the quality of a frozen density -- generated without taking into account the response of an active subsystem -- which consequently compromises the quality of the embedding potential. Secondly, no information about the response of the environment is included in the response calculations for the active subsystem.
To our knowledge, there are no studies that address the errors introduced with this approximation for 
NMR properties;  FDE studies of excitation spectra\cite{ricardi-pccp-20-26053-2018,tolle-jcp-151-174109-2019,scholz-ijqc-120-e26213-2020} and chirooptical properties\cite{niemeyer-jctc-16-3104-2020} of embedded species indicate that the presence of charge transfer or strong polarization between subsystems requires this coupling to be well described.

In this respect, the quality of the embedding potential is extremely important. Due to the use of approximate exchange-correlation and kinetic energy functionals, (ii), the embedding potential is a local (at best - semilocal) operator, which may not sufficiently well describe the redistribution of the unperturbed densities of an active subsystem and its environment upon interaction. 
Commonly used functionals for the non-additive kinetic energy potential\cite{jacob-wire-4-325-2014,jacob-jcp-126-234116-2007,banafsheh-ijqc-118-e25410-2018}
may introduce 
artificial low-energy virtual orbitals on the frozen subsystems, in consequence significantly affecting the NMR shielding calculations.\cite{jacob-jcp-126-234116-2007} 
The choice of the functionals used for the the non-additive exchange-correlation energy potential can also be 
relevant.
For instance, 
systems including transition metals may require the use of double-hybrid exchange-correlation functionals,\cite{bensberg-pccp-22-26093-2020} 
or long-range corrections.\cite{tolle-jcp-151-174109-2019}

The use of monomer basis sets, (iii), can further compromise the quality of 
results, because
it does not allow for straightforward description of charge transfer between the subsystems.
We have performed test calculations on MnO$_4^{-}$  using for each subsystem the supermolecule  basis set;
we did not pursue this line of inquiry 
because 
the small gain in accuracy did not justify the increased computational demands.
Typically, the problem is resolved by 
modifying the 
partitioning of the total molecular system into subsystems -- which 
might be a sensitive issue 
in systems involving charged complexes of transition metal atoms.
It has been observed that the inclusion of some molecules from the solvent into the active subsystem might be necessary to circumvent the errors introduced with approximations (i)-(iii).\cite{scholz-ijqc-120-e26213-2020}

\section{Conclusions}

We have employed the FDE scheme to calculate solvent shifts of the NMR shielding constants of selected transition metal nuclei in complexes in various solvents. FDE was applied at the relativistic level with the two- and four-component Hamiltonians (SO-ZORA and DC) and with the DFT method used for all subsystems.

In almost every case,  for both SO-ZORA and DC Hamiltonians, the fde(N) results are in better agreement with the supermolecular solvent shifts than the fde0 results.
Although fde(N) is an iterative
procedure,
thus 
more time consuming than fde0,  in contrast to the supermolecular approach it
still involves only calculations on separate subsystems (which can be performed in their respective basis sets in order to reduce the computational cost - as we have done in this work).
Moreover, the fde(N) procedure 
was found to converge fast (the changes in the total
energies in two last freeze-and-thaw iterations are lower than 1e-06
a.u.), thus qualitatively correct results could be obtained in smaller number of freeze-and-thaw iterations.

In comparison with recently published PCM results on the same systems,\cite{aamrpccp22} FDE values seem to be more reliable overall - the dispersion of FDE solvent shifts in the studied set is significantly lower than the one based on PCM values.
Qualitatively, both models agree in most cases 
by predicting the same signs of solvent shifts, yet neither is accurate in all studied systems. With these studies we therefore hope to encourage further tests of the FDE scheme on NMR properties, in order to understand its applicability in various complex systems and direct its further developments.

Accompanying supermolecular calculations (at the same relativistic DFT levels) allowed to assess the accuracy of the FDE approach. We found that FDE in its canonical formulation does not 
guarantee
satisfactory results, even when the role of subsystems is repeatedly interchanged in the freeze-and-thaw procedure leading to  self-consistently optimized electron densities of subsystems.
The electronic contributions to solvent shifts cannot be neglected, even when a significant portion of the solvent shifts is actually recovered by mechanical solvent effects -- the latter governed by the relaxation of subsystems' geometries in 
the formation of the supermolecule.
This
calls for a better description and analysis of these electronic effects on the shifts, one that goes beyond the 
fde(N) approach. 
Another interesting dilemma are the differences between solvent shifts at the SO-ZORA and DC levels, which emphasize that one should not rely on the cancellation of errors when calculating solvent shifts to the same degree as in the evaluation of the NMR chemical shifts.
Recalling that NMR properties depend on the magnetic-field induced currents, which may encompass the whole supermolecule, a "coupled-FDE" approach 
in which each subsystem interacts with the other, 
with both simultaneously perturbed 
by the external field, appears to be 
an extension worth exploring.
Considering the number of approximations in our work (such as the use of DFT and monomer basis sets)
the improved performance of fde(N) with respect to fde0 successfully confirms the advantages of the FDE approach.

\section*{Acknowledgements}
MO would like to thank Andr\' e Severo Pereira Gomes for fruitful discussions and for his comments on the draft of this article.

\section*{Research Resources}
This research was supported in part by PLGrid Infrastructure.

\section*{Conflict of interest}
There are no conflicts to declare.

\section*{Data Accessibility}
The data that support the findings of this study (inputs and outputs of calculations) are openly available in Zenodo at DOI: \href{https://doi.org/10.5281/zenodo.4883729}{10.5281/zenodo.4883729}

\end{document}